# Construction d'une plateforme intégrée pour l'analyse spatiale des inégalités d'exposition environnementales : résultats en Picardie et Nord-Pas-de-Calais

J. CAUDEVILLE (1,2), G. GOVAERT (2), C. BOUDET (1), R. BONNARD (1), S. DENYS (1), A. CICOLELLA (1)
(1) Institut National de l'Environnement Industriel et des Risques (INERIS)
(2) UMR 6599, Université de Technologie de Compiègne (UTC)

- *Introduction/Objectif*

L'analyse du lien entre l'environnement et la santé est devenue une préoccupation majeure de santé publique comme en témoigne l'émergence des deux Plans Nationaux Santé Environnement. Pour ce faire, les décideurs sont confrontés au besoin de développement d'outils nécessaires à l'identification des zones géographiques pour lesquelles on observe une surexposition à des substances toxiques. L'objectif du projet SIGFRIED 1 est de construire un indicateur spatialisé permettant d'évaluer l'exposition de la population française aux substances chimiques et ses déterminants.

- *Matériels/Méthodes*

La construction de l'indicateur spatialisé repose sur le couplage de deux approches que sont l'évaluation des expositions et la spatialisation des données. Une plateforme de modélisation est construite présentant les caractéristiques suivantes :
- une approche intégrée prenant en compte la variété des situations de transfert des polluants dans les compartiments environnementaux et les média d'exposition aux différentes échelles spatiales (globale, régionale et locale) et temporelles ;
- la possibilité de décrire sur une échelle fine les principales sources polluantes, compartiments environnementaux (eau, air, sol), voies d'exposition (inhalation, ingestion) et les populations résidentes des zones évaluées ;

L'évaluation des expositions est réalisée par le biais d'une modélisation multimédia. Les problèmes épistémologiques liés à l'absence de données sont palliés par la mise en œuvre d'outils utilisant les techniques d'analyse spatiale.

- *Résultats*

Le calcul de l'exposition est réalisé sur une durée de 30 ans sur une maille de 1 km de côté.
Un exemple est fourni sur la Région Nord-Pas-de-Calais et Picardie, pour les particules en suspension ($PM_{10}$ et $PM_{2.5}$), le cadmium, le chrome, le nickel et le plomb. Des méthodes de détection de clusters ont été utilisées pour localiser les zones de surexposition potentielle. Pour le Nord-Pas-de-Calais, l'indicateur permet de définir deux zones pour le cadmium et trois zones pour le plomb. Celles-ci sont liées à l'historique industriel du Nord-Pas-de-Calais : le bassin minier, les activités métallurgiques et l'agglomération lilloise. La contribution des différentes voies d'exposition varie sensiblement d'un polluant à l'autre.

- *Discussion/Conclusion*

Ces résultats présentent de fortes incertitudes liées à de nombreux biais mais permettent de mettre en exergue des zones de fortes contaminations. Les cartes d'exposition obtenues seront par ailleurs intégrées dans le projet CIRCE (Cancer Inégalités Régionales Cantonales et Environnement) pour être confrontées aux cartes de disparité de répartition de mortalité par cancer et d'indicateur socio-économique, construites par ailleurs par les Observatoires Régionaux de la Santé (ORS) des régions partenaires du projet. Le SIG ainsi construit constitue la base d'une plateforme où les données d'émission à la source, de mesures environnementales, d'exposition et de santé sont associées.

Mots-clés : plateforme, SIG, indicateur, exposition, environnement, modélisation, inégalité.

## LES ANNEXES

Le projet CIRCE a été élaboré en 2004 avec pour objectif de quantifier l'exposition humaine aux polluants environnementaux cancérigènes, de saisir ses hétérogénéités spatiales et d'évaluer son influence sur le risque de cancer. Après la Picardie, l'Ile-de-France, le Nord-Pas-de-Calais et Rhône-Alpes, la Bretagne a rejoint le groupe des ORS. La contribution des ORS est de fournir les cartes de mortalité par cancer. Il repose sur un partenariat entre l'INERIS, l'INSERM et les ORS Picardie, Ile-de-France, Nord-Pas-de-Calais Rhône-Alpes et Bretagne.

### 1) Description de la plateforme d'intégration et de modélisation

**La plateforme de modélisation** développée permet le couplage des données de sources polluantes, de qualités environnementales, des média d'exposition, des doses d'exposition et des effets sanitaires. Une base de données est construite permettant la description sur une grille d'étude de maille kilométrique de chacun des éléments de la chaine source-effet des paramètres pertinents de l'étude:
- les données d'émission atmosphériques par la spatialisation d'un cadastre d'émission,
- les niveaux environnementaux par la construction d'indicateur à partir des données environnementales eau-air-sol,
- les concentrations dans les média d'exposition et les doses d'exposition à partir de la modélisation multimédia présentée plus loin.

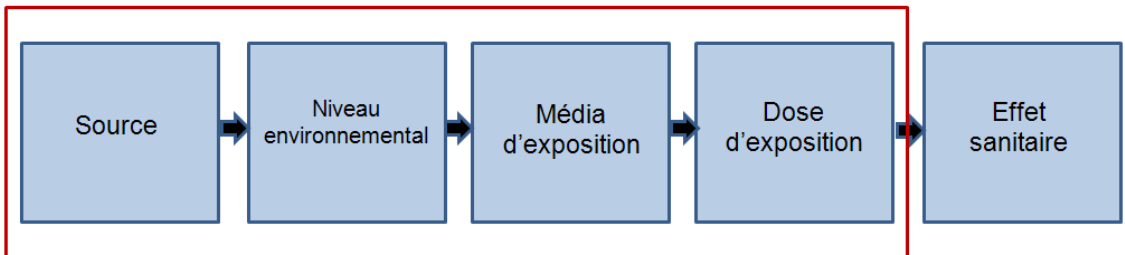

Figure 1 : Schéma conceptuel de la chaîne source-effet.

La plateforme est construite et paramétrée par un SIG qui permet la collecte, la gestion, la manipulation, l'analyse, la modélisation et l'affichage des données à référence spatiale.
Le caractère multi-échelle du système permet la gestion d'études fines sur le territoire français, mettant en évidence les zones à hauts niveaux d'émissions, de fortes contaminations des compartiments environnementaux ou de surexposition potentielle. Les données de niveau environnemental sont par ailleurs utilisées comme données d'entrée au modèle d'exposition multimédia pour l'estimation des doses d'exposition.par l'intermédiaire du calcul des concentrations dans les médias d'exposition.
Un modèle multimédia d'exposition est construit pour permettre le calcul des doses d'exposition de populations cibles liées à l'ingestion de produits d'alimentation, d'eau de consommation, de sol et à l'inhalation de contaminants atmosphériques.

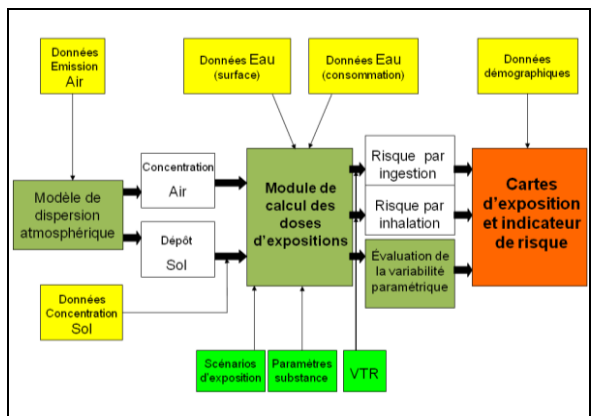

Le modèle (figure 2) est alimenté par des bases de données géoréférencées de différents types : environnemental (eau, air, sol, alimentation), comportemental, démographique et interfacé avec un Système d'Information Géographique (SIG) au sein de la plateforme.
Le modèle utilise les équations de transfert de polluant de la source à l'individu. Cet indicateur correspond à l'exposition des populations sur une durée de 30 ans dans le cas d'une non-modification des pratiques polluantes ou de mise en place de politique de restauration des milieux.

Figure 2 : Description du module de calcul de l'indicateur d'exposition

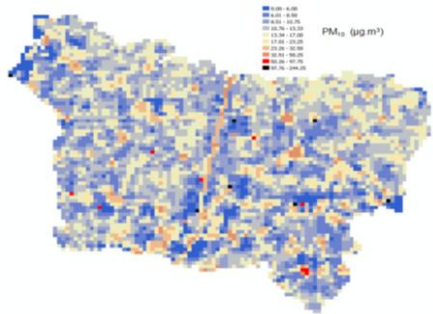

### 2) Exemple de résultats obtenus

Un SIG est utilisé pour cartographier l'exposition évaluée par polluant sur la région Nord-Pas-de-Calais et Picardie. Deux cartes sont présentées dans les figures 3 et 4 : la première correspondant aux concentrations dans l'air en $PM_{10}$ en Picardie, la deuxième aux expositions de plomb en Nord-Pas de Calais.

Figure 3 : Cartographie des concentrations en $PM_{10}$ en Picardie (en $\mu g.m^{-3}$).

Pour le Nord-Pas-de-Calais, cinq catégories sont définies allant de A à E selon le niveau d'exposition. L'indicateur correspondant au plomb permet de révéler trois zones de surexposition potentielle parmi lesquelles les grandes sources de pollution bien identifiées en Nord-Pas-de-Calais (au centre dans le bassin minier : Metaleurop et Umicore ; à l'est : Mortagne-du-Nord) et une troisième zone correspondant à l'agglomération lilloise, caractérisée par une pollution locale induite par te trafic automobile et les rejets industriels auxquels s'ajoute un apport industriel extérieur originaire du bassin minier et du littoral. La figure 4 présente la contribution des différentes voies d'exposition au calcul de la DJE ($mg/kg^{-1}.j^{-1}$, échelle logarithmique) du plomb. Le tableau rassemblent l'ensemble des résultats des mailles de l'étude et permettent d'estimer le poids des parts locales et ubiquitaires. Ici l'effet local est prépondérant. L'ingestion de viande, de lait, d'eau de consommation, de fruits et de légumes sont les contributeurs prépondérants au calcul de la DJE. Des méthodes statistiques (Moran's I test) sont utilisées pour identifier des zones de surexposition potentielle significativement élevée.

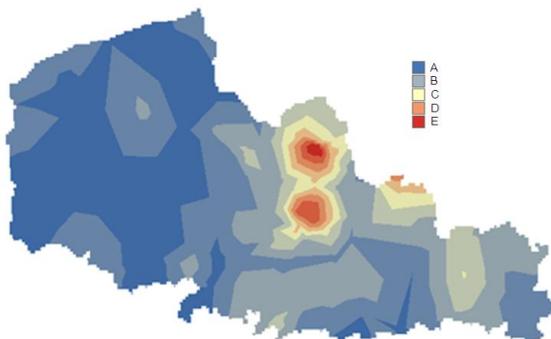

Figure 4 : Cartographie de l'indicateur d'exposition pour le plomb en Nord-Pas-de-Calais

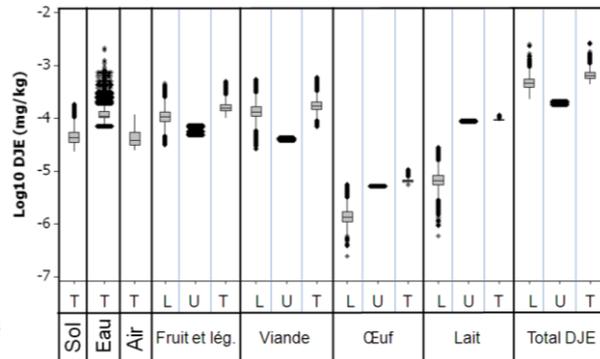

Figure 5 : Contribution des différentes voies d'exposition au calcul de la DJE. (L : local ; U : ubiquitaire et T : total).

### 3) Publication et communication

J. Caudeville, G. Govaert, R. Bonnard, O. Blanchard, A. Ung, B. Bessagnet, A. Cicolella, construction d'un indicateur d'exposition spatialisé de l'environnement : application au Nord-Pas-de-Calais, Air Pur 76 (2009), p49-55. **Article**

Construction d'un indicateur d'exposition spatialisé de l'environnement. Intégration de bases de données environnementales dans un système d'information géographique. CAUDEVILLE J., GOVAERT G., BONNARD R., BLANCHARD O., CICOLELLA A.   01/01/2009   Actes de la conférence "STIC et environnement ", 16-18 juin 2009, Calais. **Présentation orale (article associé)**

Projet SIGFRIED 1 : SIG Facteurs de Risques Environnementaux et Décès par cancer. Intégration de bases de données environnementales dans un SIG pour servir à l'analyse des disparités géographiques de cancer. CAUDEVILLE J., MASSON J.B.   01/01/2008   Actes des journées interdisciplinaires de la qualité de l'air, 7-8 février 2008, Lille. **Présentation orale (article associé)**

Using geographic information systems to build an indicator of human exposure to local environmental quality concerning chronic health risks. CAUDEVILLE J., GOVAERT G., BLANCHARD O., BONNARD R., CICOLELLA A., Proceedings of the 6th International conference on innovations in exposure assessment, 17-20 august 2009, Boston, USA. **Présentation orale**

Assessing heavy metal exposure from ingestion with multimedia exposure model operating in a Geographic Information System environment. CAUDEVILLE J., GOVAERT G., BLANCHARD O., BONNARD R., CICOLELLA A. Abstract book of the ISES 2009 Annual conference "Transforming exposure science in the 21st century", 1-5 november 2009, Minneapolis, MN, USA. **Présentation orale du poster**